# Crystal growth with oxygen partial pressure of the $BaCuSi_2O_6$ and $Ba_{1-x}Sr_xCuSi_2O_6$ spin dimer compounds

Natalija van Well[1,2,*], Pascal Puphal[1], Björn Wehinger[2,3], Mariusz Kubus[2,4], Jürg Schefer[2], Christian Rüegg[2,3], Franz Ritter[1], Cornelius Krellner[1], Wolf Assmus[1]

[1] Physikalisches Institut, Goethe-Universität Frankfurt, D-60438 Frankfurt am Main, Germany
[2] Laboratory for Neutron Scattering and Imaging, Paul Scherrer Institute, CH-5232 Villigen, Switzerland
[3] Department for Quantum Matter Physics, University of Geneva, CH-1211 Geneva, Switzerland
[4] Department of Chemistry and Biochemistry University of Bern, CH-3012 Bern, Switzerland

ABSTRACT

$BaCuSi_2O_6$ is a quasi-two dimensional spin dimer system and a model material for studying Bose-Einstein condensation (BEC) of magnons in high magnetic fields. The new $Ba_{1-x}Sr_xCuSi_2O_6$ mixed system, which can be grown with x ≤ 0.3, and $BaCuSi_2O_6$, both grown by using a crystal growth method with enhanced oxygen partial pressure, have the same tetragonal structure ($I4_1/acd$) at room temperature. The mixed system shows no structural phase transition, so that the tetragonal structure is stable down to low temperatures. The oxygen partial pressure acts as control parameter for the growth process. A detailed understanding of the crystal structure depending on the oxygen content will enable the study of the spin dynamics of field-induced order states in this model magnetic compound of high current interest with only one type of dimer layers, which shows the same distance between the Cu atoms, in the structure.

## I. INTRODUCTION

$BaCuSi_2O_6$ is known as Han purple, a pigment already used by Chinese artists centuries ago[1,2,3]. Interestingly, studies have shown that it is a quasi-two dimensional spin dimer system. The material is observed to have a singlet ground state in zero magnetic field with a large gap to the lowest excited triplet states[4,5]. This observation is consistent with the dimerized structure of the compound and dominant antiferromagnetic exchange in the dimers. The spin gap closes completely with a magnetic field of around 23 T, so that cooling in a large applied field results in a state, which is characterized by long-range magnetic order[6,7]. Hence, a quantum critical point at around 23 T and T=0 K separates the quantum paramagnetic regime from the ordered state.

The preparation of the $BaCuSi_2O_6$ and $BaCuSi_4O_{10}$ pigments at different temperatures in air has been done using a mixture of $BaCO_3/BaSO_4$-CuO-$SiO_2$ of different stoichiometry with $BaCO_3$ and $BaSO_4$ being applied as starting substances for this mixture[8,9]. The results show that temperature is very important for the phase building process of polycrystalline samples. Using $BaCO_3$, CuO and $SiO_2$ with a molar proportion of 1 : 1 : 2 at a temperature of around 1000°C the following phases were formed: $BaCu_2Si_2O_7$, $BaCuSi_2O_6$ and a small amount of $BaCuSi_4O_{10}$. At a high temperature of around 1100°C the phase $BaCuSi_2O_6$ was found, but at this temperature a decomposition of the compound is observed instead of melting. The reduction of $Cu^{2+}$ to $Cu^{1+}$ occurs at temperatures above 1050°C. Using $BaSO_4$ and molar proportions 1 : 1 : 2 at temperatures of 1000°C and 1100°C results in a phase building of $BaCuSi_2O_6$ and $BaCuSi_4O_{10}$. A model of progressive silicate condensation has been developed to describe the formation of these compounds. Starting from the basic structural element $SiO_4^{4-}$ (orthosilicate), the $Si_2O_7^{6-}$ ion (disilicate) and, in the next step, the four-ring unit $Si_4O_{12}^{8-}$ (cyclotetra-silicate) are generated by increasing condensation. This forms the basic silicate unit of Han Purple. $Si_4O_{12}^{8-}$ may condense and build connected layers of puckered $Si_8O_{20}$ eight-membered ring units and planar four-membered rings. The parameter of building different silicate formations in air depends on the temperature range.

Single crystals of this material were prepared using the floating zone[6] or the slow cooling flux techniques[7]. The floating zone growth of this material was carried out in an $O_2$ flow with varied speed of the growth. In the first run the speed of the growth was set to 50 mm/h, and in the second run to 0.5 mm/h. In comparison the crystal growth using the slow cooling flux technique with $LiBO_2$ was carried out in air atmosphere.

The crystal structure of $BaCuSi_2O_6$ at room temperature is tetragonal ($I4_1/acd$) and establishes $CuO_4$ layers, which are retained on their edges with $SiO_4$ tetrahedra[10]. The layers are linked in pairs such that the distance between two Cu atoms is only 2.75 Å. These bi-layers form well-defined structural and magnetic dimers and are separated from each other through $Ba^{2+}$ cations. Each spin ½ $Cu^{2+}$ ion interacts with the four $Cu^{2+}$ ons of the next layer. The interlayer coupling of the bilayers is antiferromagnetic (below 1K) and a perfect frustration may be realized[7]. Recently, an additional structural phase transition of the $BaCuSi_2O_6$ system was discovered at around 100K. Its structure changes to orthorhombic ($I$bam)[11] and, furthermore, an incommensurable structure was observed at temperature below 16K[12]. Thereby, also the $CuO_4$ layers split into a minimum of two types of neighboring double layers. These double layers show different couplings within the dimers. This was confirmed using inelastic neutron scattering (INS)[5] and nuclear magnetic resonance (NMR)[13]. Nevertheless, all experimental and theoretical studies assume that the magnetic frustration is an integral part of the $BaCuSi_2O_6$ system. The influence of the structural transitions on the quantum phase transition of Bose-Einstein type is essential to understand the importance of geometrical frustration for the unique dimensional reduction, reported for the first time in $BaCuSi_2O$ [7] and thereafter discussed in a large number of related theoretical studies[14,15,16]. To verify this theory, we used a method with oxygen partial pressure for the crystal growth of $BaCuSi_2O_6$ and the $Ba_{1-x}Sr_xCuSi_2O_6$ mixed system, which has to be recognized for the examination of the Bose-Einstein condensation.

In this paper we present the results of the synthesis and the phase characterization of the polycrystalline samples of $BaCuSi_2O_6$ and the $Ba_{1-x}Sr_xCuSi_2O_6$ mixed system in Sec. III. In Sec. IV we provide experimental details of the crystal growth with different methods and the structural characterization. The structure investigation at low temperature of BaCuSi$_2$O$_6$ and the Ba$_{1-x}$Sr$_x$CuSi$_2$O$_6$ mixed system is presented in Sec. V, followed by a conclusion.

## II. EXPERIMENTAL DETAILS

The powder x-ray diffraction measurements (PXRD) were conducted with a Bruker D8-Focus diffractometer and for the low-temperature diffraction with a Siemens D-500 diffractometer with CuK$\alpha$ radiation ($\lambda$=1.5406 Å). Cooling of the powder was performed with a closed-cycle helium refrigerator. Unit cell parameters were refined with the GSAS Suite of Rietveld program using the general structure analysis software packet and the EXPGUI program[18]. The thermal stability of these crystals was investigated by means of a differential scanning calorimetry (DSC) and the thermogravimetry (TG) by using a Netzsch STA 409 System. The crystal growth with oxygen partial pressure was performed in a furnace, which has a SiC-heating element, and can be used up to 1450°C. For the oxygen partial pressure an Al$_2$O$_3$ pipe was utilized, which was placed in the SiC heating element. The experiments were carried out with oxygen pressure of up to 5 bar. The chemical compositional analysis of the samples was conducted via energy dispersive X-ray analysis (EDX) using Zeiss DSM 940A. X-ray Laue back-reflection photographs were taken with imaging-plate (IP)-type detector and a tungsten tube operated at 14 kV, 20 mA. The exposure time for a back-reflection Laue photograph was 10-15 min. For the conoscopic image the Zeiss microscope "Universal" was used.

## III. SYNTHESIS AND PHASE CHARACTERIZATION

The synthesis of BaCuSi$_2$O$_6$ is based on the reaction

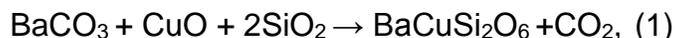
$$BaCO_3 + CuO + 2SiO_2 \rightarrow BaCuSi_2O_6 + CO_2, \quad (1)$$

with molar proportions of the components 1 : 1 : 2. As there is a variety of reported temperatures and durations of the solid state reaction[1,2,17], it is necessary to establish the optimized temperature for the synthesis. We have carried out a number of investigations, where sample preparations were done by different multistage sintering processes, lasting up to three months. Fig. 1 a) shows three of our samples, which were sintered at temperatures between 1030°C and 1035°C. The distinction between the samples can be seen, as they show a variety of blue colors, whereby illustration (iii) shows a darker blue color, which corresponds to the highest sinter temperature. Purity and crystallinity was examined by x-ray powder diffraction. Small impurities are always present in this compound and may not be detected using powder diffraction with standard laboratory equipment. However, they can be determined by synchrotron x-ray diffraction. We structurally classify the materials according to our PXRD data, see Fig. 1b). The selected Bragg reflections show the improvement of the crystallinity as a result from only a small change of the sintering temperature. A sintering temperature of 1030°C leads to only a small amount of impurity phases (Fig. 1c), whereas the compound, sintered at a 1035°C, shows a pure $BaCuSi_2O_6$ phase with higher Bragg intensities at higher scattering angle ranges, indicating an improved crystallinity. The duration of sintering was 2 weeks for both temperatures.

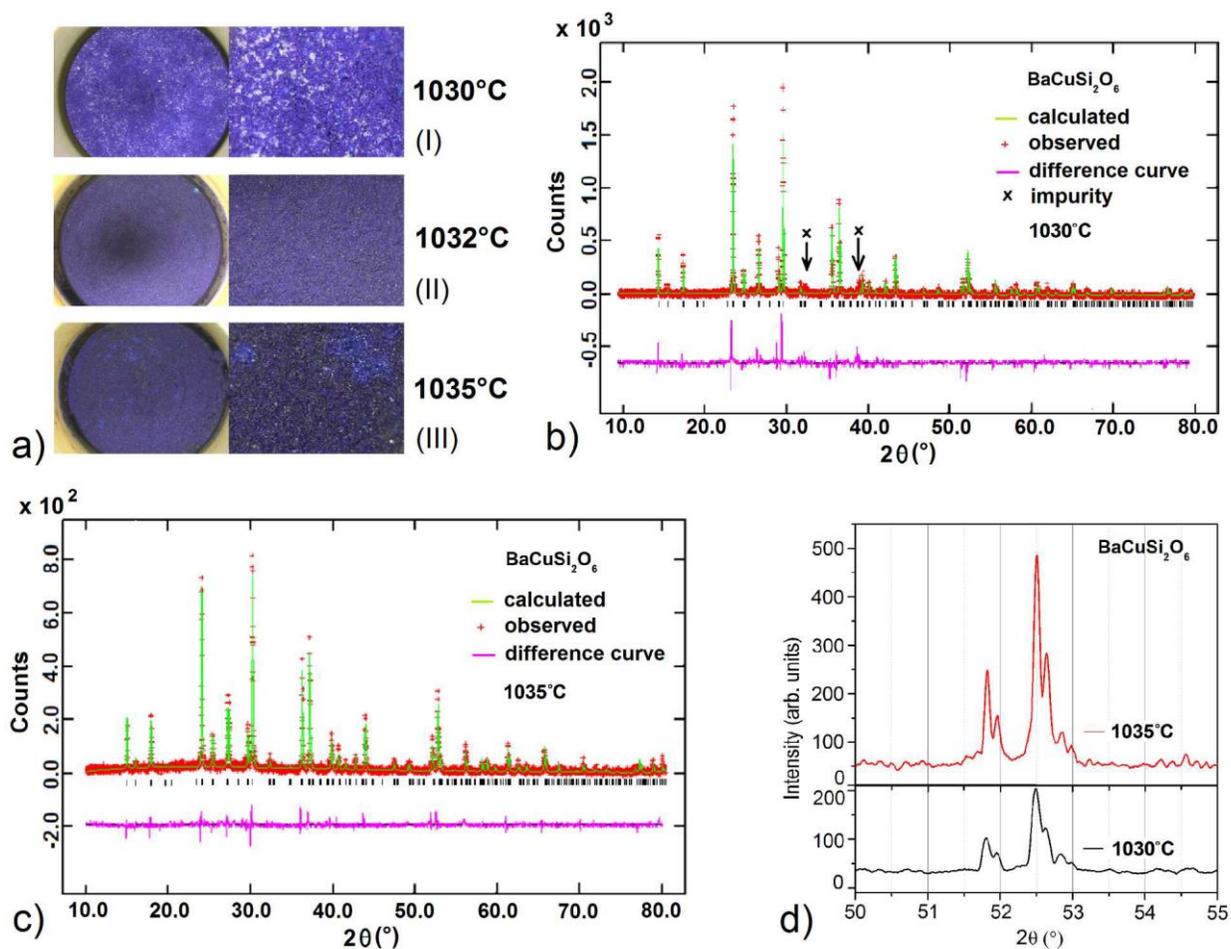

Figure 1: a) $BaCO_3$-$CuO$-$2SiO_2$ mixture synthesized at a variety of temperatures. Refined PXRD data at room temperature, b) and c) using the GSAS Suite of Rietveld program[18], for two powder samples prepared at b) 1030°C and c) 1035°C, which both show the tetragonal structure type $I4_1/acd$; d) comparison of both powder samples for a selected angle range

The lattice parameters of polycrystalline $BaCuSi_2O_6$, synthesized at 1030°C and 1035°C, are given in Tab. 1. They have both the same space group ($I4_1/acd$) at room temperature with practically identical cell parameters. As the temperature has a big influence on the phase building process, it is worth noting that the structure details of both refined PXRD data are very similar. The atomic positions of this compound match the published data at 293K[10]. As there are only a few different impurity phases (see Fig. 1b)), they can be neglected in the refinement.

Table 1: Lattice parameters and atomic positions at room temperature for the polycrystalline samples of $BaCuSi_2O_6$ prepared at 1030°C and 1035°C, refined from PXRD data

|  | $BaCuSi_2O_6$ - 1030°C | | | $BaCuSi_2O_6$ - 1035°C | | |
| --- | --- | --- | --- | --- | --- | --- |
| $a = b$ (Å) | 9.971(1) | | | 9.970(1) | | |
| $c$ (Å) | 22.304(4) | | | 22.304(4) | | |
| Atom | x | Y | z | x | y | z |
| Ba 16e | 0.238(1) | 0 | 0.25 | 0.245(4) | 0 | 0.25 |
| Cu 16d | 0 | 0.25 | 0.062(1) | 0 | 0.25 | 0.062(1) |
| Si 32g | 0.281(1) | 0.758(1) | 0.876(2) | 0.276(1) | 0.756(2) | 0.872(1) |
| O1 32g | 0.177(1) | 0.721(3) | 0.823(1) | 0.208(2) | 0.743(6) | 0.810(1) |
| O2 32g | 0.386(2) | 0.882(2) | 0.865(1) | 0.366(2) | 0.851(2) | 0.861(1) |
| O3 32g | 0.311(2) | 0.788(2) | 0.071(1) | 0.310(1) | 0.781(3) | 0.065(1) |
| $X^2_{red}$ | 1.803 | | | 1.523 | | |
| wRp | 0.1927 | | | 0.1713 | | |
| Rp | 0.1523 | | | 0.1353 | | |

The possibility to reduce the temperature required to form the polycrystalline phase of $BaCuSi_2O_6$ was investigated by adding flux ($Na_2CO_3$, MgO, CaO and $Al_2O_3$) before sintering the ingredients, whereby different flux compositions were used, which amounts to less than 10% of the total mass in respect to $BaCO_3$, CuO and $SiO_2$, respectively. Without taking into account small impurities, which are always present in this compound, a pure phase of $BaCuSi_2O_6$ was reached at a sintering temperature of 820°C by using the composition 0.5 $Na_2CO_3$, 0.2 MgO, 0.1 CaO and 0.2 $Al_2O_3$. This implies a reduction of the sintering temperature by almost 200°C. In addition, the sintering time could also be reduced, depending on the composition of the flux. The multistage process of sintering this material generally lasts up to one month. In our case, by using flux, the sintering period was reduced to less than five days.

In order to investigate the preconditions for forming the $BaCuSi_2O_6$ phase, a differential scanning calorimetry/thermogravimetry (DSC/TG) was performed. Such measurements of $BaCO_3$-CuO-$2SiO_2$ were conducted with a pre-defined temperature profile in three steps by using an $Al_2O_3$ crucible. The temperature profile consists of 3 segments in a pure oxygen atmosphere of 0.3 bar, with the highest temperature of 1105°C. In the first dynamic segment the compound was heated up to 1105°C by using a heating rate of 5°C per minute. The compound rests in an isothermal segment for 30 minutes thereafter, before being cooled down to room temperature in the second dynamic segment by using a cooling rate of 5°C per minute. Fig. 2 shows the first dynamic segment. Two peaks in the DSC-curve were noticed at 573°C and 816°C. The first peak indicates a structural phase transition from trigonal α-quartz into hexagonal β-quartz[19]. The second peak at 816°C can be explained with the transformation of $BaCO_3$ from the rhombic into the hexagonal crystal structure. The temperature of the phase transition is in excellent agreement with the reported 813°C[20]. The third peak at 950°C, which consists of two parts, is characterized by an overlap of the endothermic and the exothermic reaction. The first part represents the decay of carbonate ($BaCO_3$), which is supposed to happen at around 1360°C, but can be much lower in the presence of other materials, like $SiO_2$ in our case[19]. The first derivative of the TG-curve reveals the peak position to ca. 950°C. This belongs to the decay of carbonate. The second part of the third peak (continuous line at ca. 950°C in Fig. 2a)) marks the starting point of the actual phase building process of $BaCuSi_2O_6$.

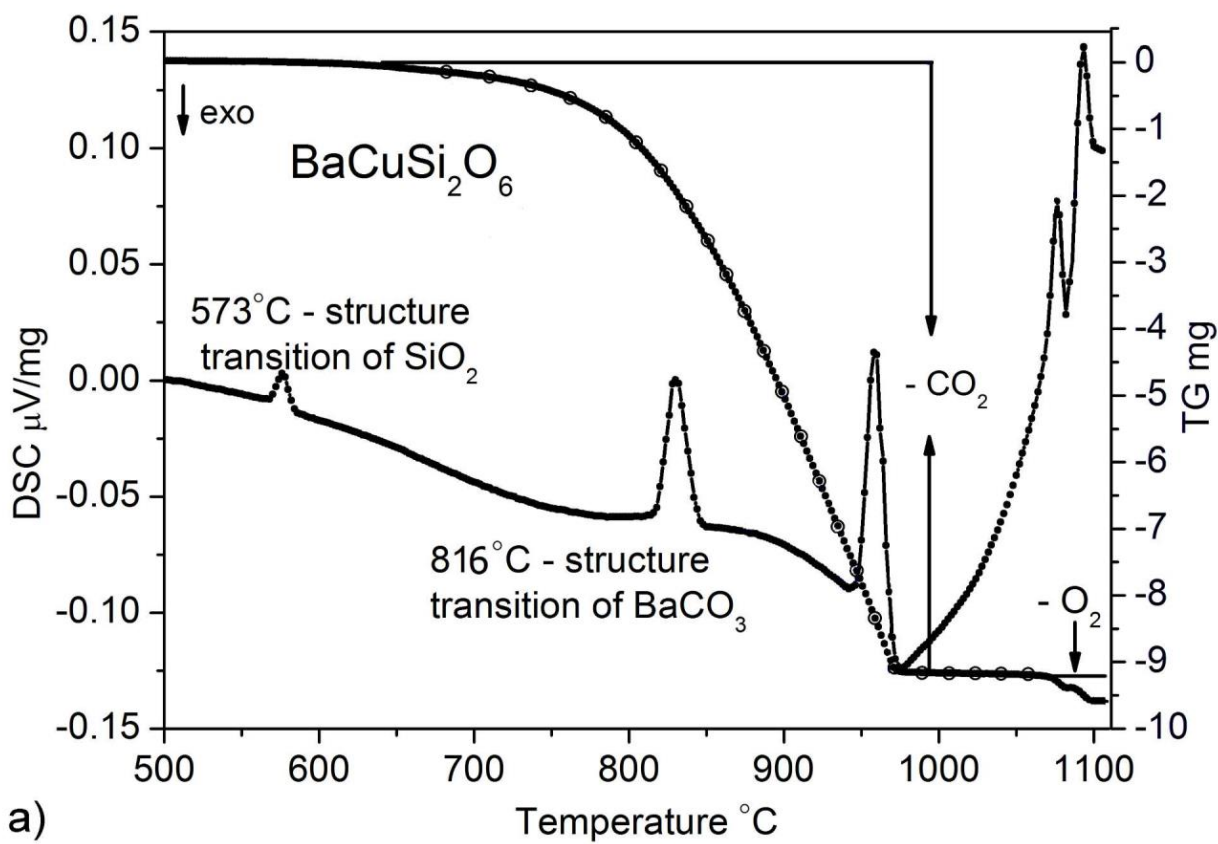

a)

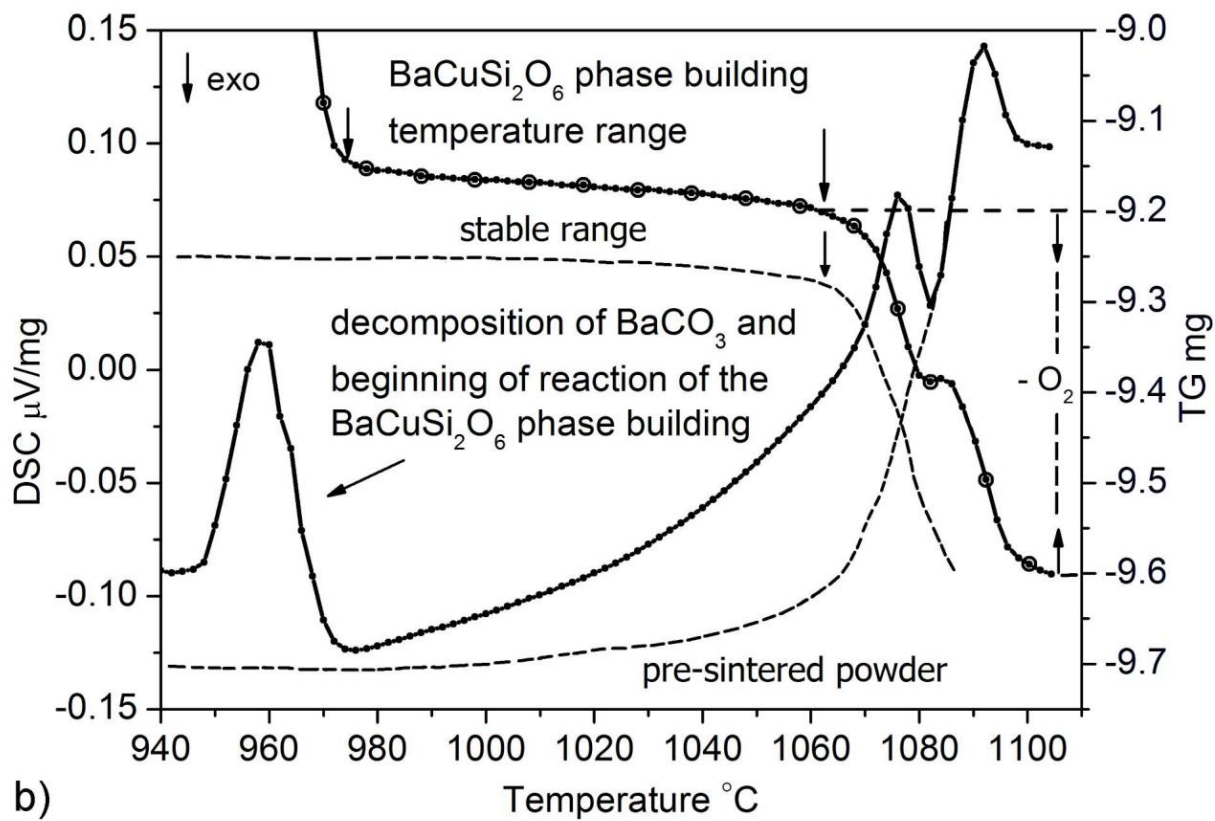

Figure 2: a) DSC/TG-curve of a $BaCO_3$-$CuO$-$2SiO_2$ mixture heated from 250°C up to 1105°C for the synthesis of $BaCuSi_2O_6$ in an oxygen atmosphere of 0.3 bar, the sample weight was 84.21 mg, the heating rate 5 K/min; b) detailed view of the $BaCuSi_2O_6$ phase building temperature range, whereby the reduction of $Cu^{2+}$ to $Cu^{1+}$ in the continuous line with circles occurs at a temperature above 1065°C in an oxygen atmosphere of 0.3 bar; the dashed line shows the stable range of the $BaCuSi_2O_6$ phase (pre-sintered powder) with a reduction of $Cu^{2+}$ to $Cu^{1+}$ at the same temperature of above 1065°C in an oxygen atmosphere of 0.3 bar

The loss in mass during the first part of the TG-curve (see continuous line with circles in Fig. 2 a)) is equal to 10.9% and can be attributed to the decay of carbonate. According to theoretical considerations the loss of mass resulting from a total decay of $BaCO_3 \rightarrow BaO+CO_2$ should equal to 11.1%, which is in excellent agreement with our observation. The resolution limit of the TG measurements is 2 µg and for DSC less than 1µV.

Between 965°C and 1065°C the TG signal is almost constant (see Fig. 2b). In this temperature range $BaCuSi_2O_6$ builds up at a maximum oxygen partial pressure of 0.3 bar (above ambient pressure) in our apparatus. Other authors report a range between 950°C and 1050°C in air atmosphere[8]. The temperature for building the phase thus increases at higher oxygen partial pressure. It can be assumed that this correlation between temperature and oxygen partial pressure only exists until the optimal oxygen partial pressure is reached. A further increase above this optimal oxygen partial pressure will lead to a decay of the material, because of its low chemical stability[9]. The fourth and fifth peaks of the DSC-curve starting at 1065°C display the reduction of $Cu^{2+}$ to $Cu^{1+}$, as outlined in Fig. 2b). This effect is clearly demonstrated in the form of a step, which starts later. Depending on the sample, in this case on a pellet from the $BaCuSi_2O_6$ mixture, the loss of oxygen at this measurement occurs in two steps. The additional loss in mass agrees with reported values[20] for temperature above 1050°C in air atmosphere:

$$3BaCuSi_2O_6 \xrightarrow{>1050C} BaCuSi_4O_{10} + 2BaSiO_3 + Cu_2O + \tfrac{1}{2}O_2. \quad (2)$$

We observe a loss in mass equal to 0.6%, which corresponds to the loss of oxygen as a consequence of the partial reduction of $Cu^{2+}$ to $Cu^{1+}$. In fact, the total decomposition would lead to a loss in mass of 4.03% *(2CuO → $Cu_2O$ + ½ $O_2$ )*. In Fig. 3 a comparison of two TG-curves from $BaCuSi_2O_6$ powder samples in air and in 0.3 bar oxygen partial pressure can be seen. The loss of oxygen shifts to a higher temperature and is reduced in case of applying oxygen partial pressure. Fig. 3 displays that the loss of oxygen of the powder sample occurs in one step, whereas Fig. 2b shows two steps for the pellet sample.

The pre-sintered powder is a pure $BaCuSi_2O_6$ phase, whose phase building process is concluded and which shows no peaks in DSC-curve due to increasing temperature up to 1065°C. For the pre-sintered powder (see Fig. 2b), dashed line) the reduction of $Cu^{2+}$ to $Cu^{1+}$ occurs without a step in the TG-curve and starts at the same temperature of 1065°C as for the non pre-sintered powder ($BaCO_3$-$CuO$-$2SiO_2$ mixture). The temperature for the stable range of the pre-sintered powder thus increases at higher oxygen partial pressure. Above the stable range the material will decay due to the increased temperature, whereby new phases including glass phases will be formed[8].

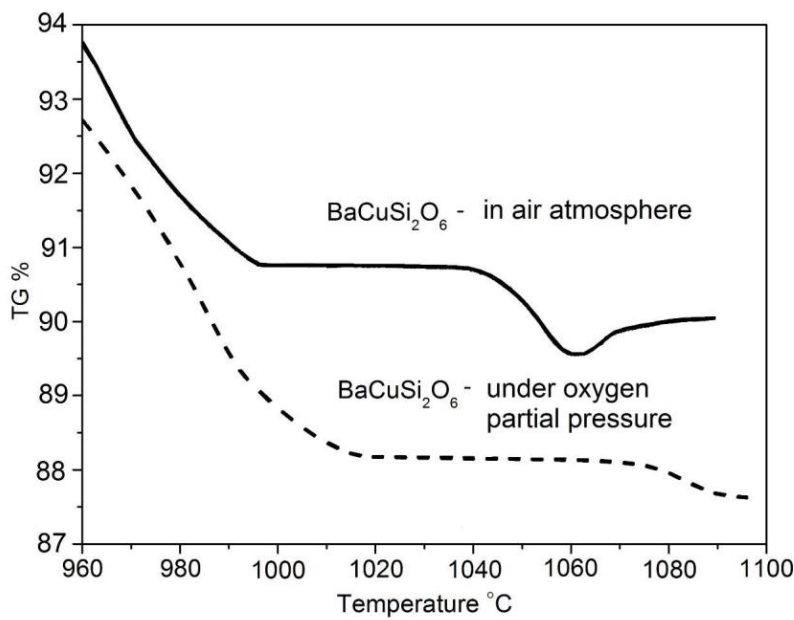

Figure 3: Comparison of two TG-curves from BaCuSi$_2$O$_6$ powder samples in air with the reduction of Cu$^{2+}$ to Cu$^{1+}$ beginning at 1040°C (continuous line), and under 0.3 bar oxygen partial pressure with the reduction of Cu$^{2+}$ to Cu$^{1+}$ beginning at 1065°C (dashed line)

The results of the DTA investigations indicate that enhanced oxygen partial pressure can be used for crystal growth of this composition. It is expected that a decay of Cu$^{2+}$ to Cu$^{1+}$ can be shifted to higher temperatures by using a specific oxygen partial pressure, which is still to be determined. Furthermore, such an increase of temperature could be sufficient to reach the melting point for this compound.

In the synthesis of the polycrystalline BaCuSi$_2$O$_6$ phase Ba could be substituted through Sr up to 30% by replacing the corresponding portion of BaCO$_3$ by SrCO$_3$ in the precursor material. It is not possible to further increase the Sr concentration without forfeiting the given structure. As it is demonstrated in Tab. 2 the synthesis temperature will be reduced with increasing Sr content.

Table 2: The chemical composition for the Sr substituted polycrystalline samples with corresponding synthesis temperatures

| Composition | Ba$_{0.95}$Sr$_{0.05}$CuSi$_2$O$_6$ | Ba$_{0.9}$Sr$_{0.1}$CuSi$_2$O$_6$ | Ba$_{0.8}$Sr$_{0.2}$CuSi$_2$O$_6$ | Ba$_{0.7}$Sr$_{0.3}$CuSi$_2$O$_6$ |
|---|---|---|---|---|
| Temperature | 1025°C | 1020°C | 1010°C | 1000°C |

The incorporation of Sr in the structure was determined by PXRD analysis. Fig. 4a) shows a comparison between BaCuSi$_2$O$_6$, Ba$_{0.9}$Sr$_{0.1}$CuSi$_2$O$_6$ and Ba$_{0.8}$Sr$_{0.2}$CuSi$_2$O$_6$. All three samples are pure Han purple phase. The detailed view (see Fig. 4b)) demonstrates that the position of the Bragg reflections for the Sr substituted compounds is shifted to higher angles. The refinement (see Table 3) confirms that the unit cells are reduced in size, accordingly.

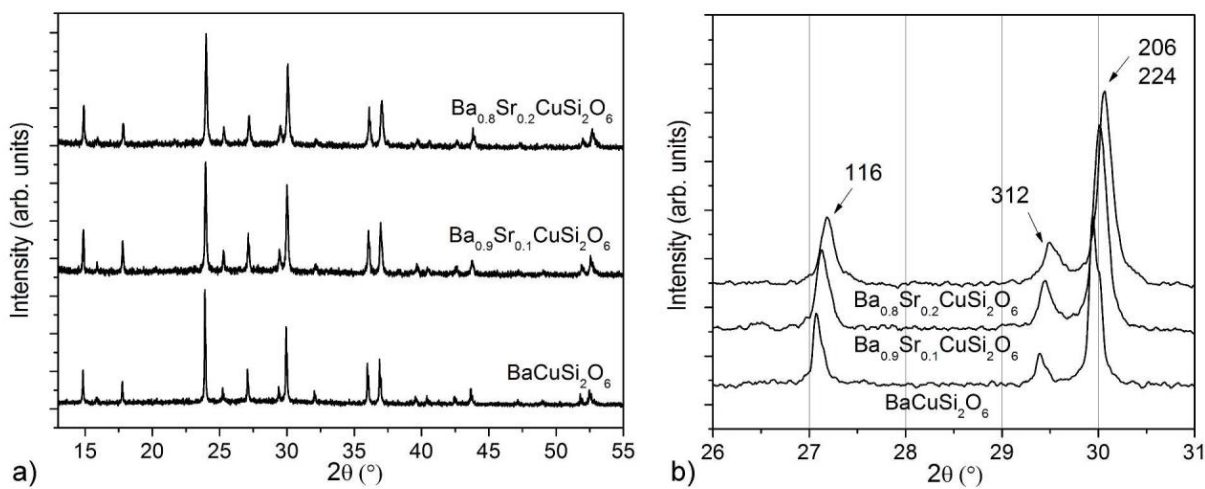

Figure 4: Comparison of the powder diffraction pattern at room temperature of the compounds Ba$_{0.9}$Sr$_{0.1}$CuSi$_2$O$_6$, Ba$_{0.8}$Sr$_{0.2}$CuSi$_2$O$_6$ and BaCuSi$_2$O$_6$: a) full angle range, and b) detailed view

The structure parameters were refined from PXRD data using the GSAS suite of Rietveld programs[18]. Tab. 3 shows the results of the lattice parameters for Sr substituted compounds. The refinement was done by using the structure model from Sparta et al.[10] for space group $I4_1/acd$. The results of the refinement show that the volume of the unit cells shrinks by 1%, in the case of increasing Sr content from 0% to 30%.

The name of the compositions given in Tab. 3 only describes the nominal values of such compounds, based on the source materials. To determine the real chemical composition, EDX and powder diffraction analysis were applied. The analysis shows that around 80% of Sr as the respective substitute was incorporated in the compounds mentioned in Tab. 2.

Table 3: Lattice constants and cell volume of $Ba_{1-x}Sr_xCuSi_2O_6$ compounds refined from PXRD data collected at room temperature

| Composition | $Ba_{0.95}Sr_{0.05}CuSi_2O_6$ | $Ba_{0.9}Sr_{0.1}CuSi_2O_6$ | $Ba_{0.8}Sr_{0.2}CuSi_2O_6$ | $Ba_{0.7}Sr_{0.3}CuSi_2O_6$ |
|---|---|---|---|---|
| a, b (Å) | 9.967(8) | 9.953(6) | 9.938(7) | 9.932(9) |
| c (Å) | 22.296(9) | 22.269(9) | 22.225(7) | 22.208(3) |
| V (Å$^3$) | 2214.909 | 2206.016 | 2195.026 | 2190.699 |

## IV. CRYSTAL GROWTH AND STRUCTURE CHARACTERIZATION

Crystal growth of $BaCuSi_2O_6$ was performed by using $LiBO_2$[7] or $KBO_2$ as flux. Both substances have similar physicochemical properties. The melting temperature of $LiBO_2$ (847-851°C) is slightly lower than that of $KBO_2$ (947°C)[21,22]. Both fluxes fulfill the requirement to dissolve a large quantity of $BaCuSi_2O_6$ powder, and, depending on the temperature, demonstrate a significant change in solubility, which is necessary to reach a supersaturation of the solution. $BaCuSi_2O_6$ crystals grow as blue rectangular bars, see Fig. 5. To grow such crystals, a Pt-crucible and a mixture of flux and $BaCuSi_2O_6$ powder with a mole ratio of 1 : 2 was used. The molten mixture of compound was cooled from 1000°C to 900°C with a cooling rate of 1°C per hour. Thereafter, the Pt-crucible was tilted in the furnace to remove the excess of flux, whereby the crystals stick to the sides of the Pt-crucible. The crystallization process of $BaCuSi_2O_6$ is due to the supersaturation of the solution during cooling. The transport of material is realized for this crystal growth method (using flux) by the solution craving up slowly on the inner wall on the side of the crucible, which finally leads to the formation of single crystals, see Fig. 5.

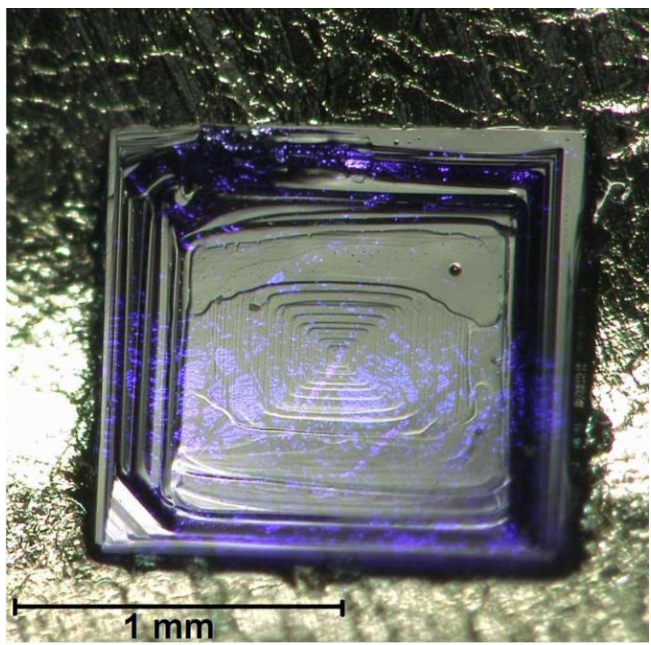

Figure 5: BaCuSi$_2$O$_6$ crystal grown in a Pt-crucible with LiBO$_2$-flux

The crystals were mechanically extracted, optionally after short etching in highly diluted HNO$_3$. An example of such an extracted crystal is shown in Fig. 6a). The polarization microscopy examination of the BaCuSi$_2$O$_6$ crystals is shown in Fig. 6b), where the same crystal is photographed in a transmitted light arrangement and in interference. The structural assignment by means of the optical image (the cross in Fig. 6b right) is an additional proof of the tetragonal structure. For growing such crystals, Pt - Au crucibles were used, where 5% Au prevents a strong adhesion of the melt to the sides of the crucible[23]. We have noticed that this applies also for the adhesion of crystals to such crucibles. Using such Pt – Au crucibles results in a reduction of the solution craving up of on the side of the crucible, and, furthermore, leads to larger single crystals, as more material is localized where crystal growth occurs.

    The examination of the chemical composition by energy dispersive x-ray analysis (EDX) showed that the real proportions of the elements comply in average with the nominal composition. Nevertheless, the percentage of oxygen varies between 60 at.% and 66 at.% and Si between 20 at.% and 22 at.%, whereas the ratio between Si and O changes from 2.8 to 3.1.

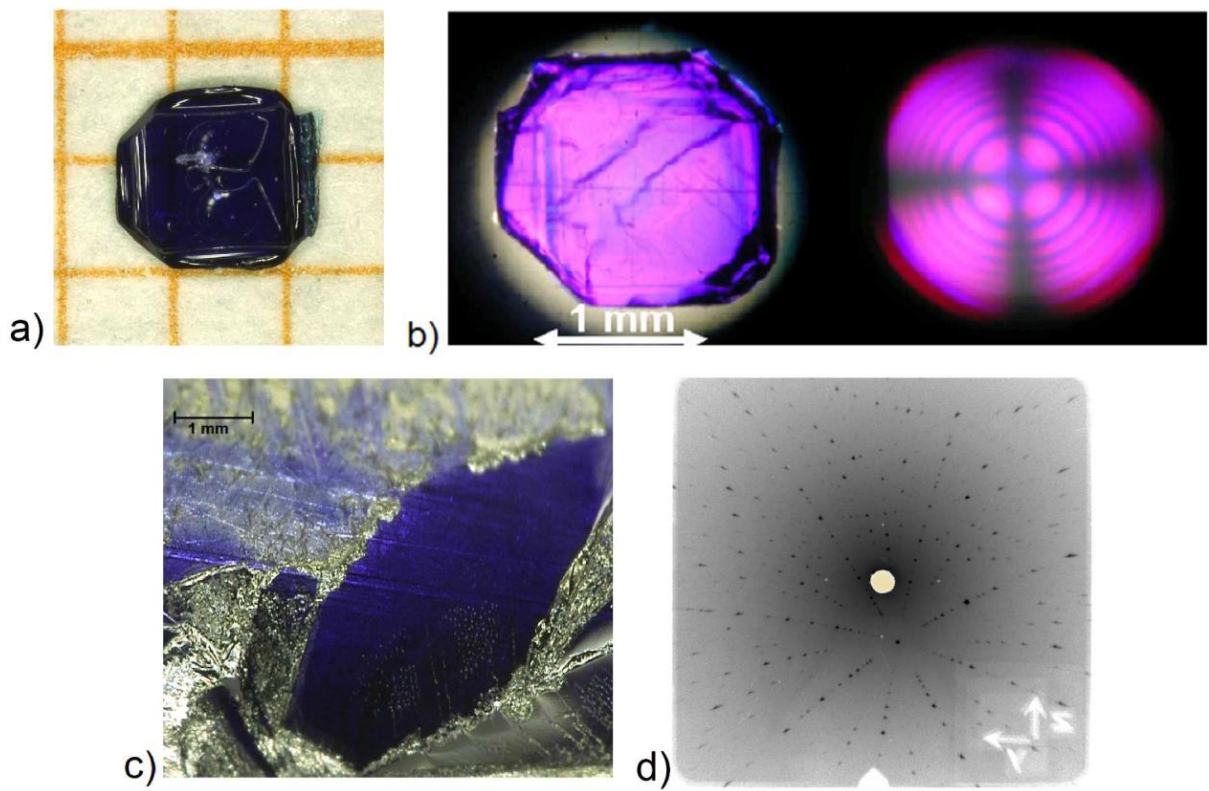

Figure 6: a) crystal of $BaCuSi_2O_6$ grown by the flux method using $LiBO_2$, b) view along c-axis and conoscopic image (right), c) crystal of $Ba_{0.7}Sr_{0.3}CuSi_2O_6$ grown under oxygen partial pressure, d) x-ray Laue-pattern viewed along c-axis with a small offset of 13°

Fig. 6c) depicts the result of crystal growth with oxygen partial pressure a monocrystalline grain of $Ba_{0.7}Sr_{0.3}CuSi_2O_6$, whereby oxygen partial pressure was used. Our investigations, using the oxygen pressure method, were done using different parameters for enhanced oxygen partial pressure (between 0.6 bar and 3 bar) and temperature (between 1120°C and 1350°C). The objective to melt the composition without reduction of $Cu^{2+}$ to $Cu^{1+}$ was realized best possible at an oxygen partial pressure of 1 bar in combination with a temperature of 1150°C, whereby oxygen partial pressure acts as a control parameter in such a way that it prevents a leakage of oxygen and therefore the decay of the composition.

The Pt-crucible with the sample was placed in an $Al_2O_3$ tube in the hot zone of the furnace (ideally without a temperature gradient over the full length of the crucible) and heated from room temperature to 1150°C over a period of 4 hours. The source material consists of a single-phase pre-sintered powder of clean $BaCuSi_2O_6$ or a powder with Sr substituted material. Without oxygen partial pressure and using a temperature of 1150°C the material decomposes in several binary and ternary phases ($BaSiO_3$, $Cu_2O$, $Ba_2Si_3O_8$). The sample rests at 1150°C for one hour before being cooled down with a cooling rate of 100°C per hour. Fig. 6d) shows the Laue-diffraction pattern with a view along c-axis of this grain, perpendicular to the surface. Good crystallinity is confirmed.

Tab. 4 presents the EDX results of the chemical composition of the single crystals of $Ba_{1-x}Sr_xCuSi_2O_6$ for x = 0.1 and 0.3, using an oxygen partial pressure for crystal growth of 0.8 bar. In consideration of the tolerance the detected values match very well to the nominal values of such compounds. The systematic error of the EDX measurement is max. 2 at.%. The analysis shows that 80% of the respective substitutes are incorporated in the grown crystals. Nevertheless, during the investigation the proportion of oxygen differs between 56 at.% and 62 at.% and Si between 20 at.% and 22 at.%, and, therefore, the ratio between Si and O changes from 2.6 to 2.8.

Table 4: Chemical composition of the Sr substituted crystals, grown with 0.8 bar oxygen partial pressure

| Crystal / Elements in at.% | Ba | Sr | Cu | Si | O |
| --- | --- | --- | --- | --- | --- |
| $Ba_{0.9}Sr_{0.1}CuSi_2O_6$ | 9.97 | 0.86 | 9.60 | 21.40 | 58.17 |
| $Ba_{0.8}Sr_{0.2}CuSi_2O_6$ | 9.53 | 1.67 | 10.21 | 20.50 | 58.11 |
| $Ba_{0.7}Sr_{0.3}CuSi_2O_6$ | 8.98 | 2.29 | 10.40 | 21.64 | 56.69 |

The investigation of the chemical composition of the single crystals of $BaCuSi_2O_6$, grown under oxygen pressure, was also conducted with EDX. The oxygen content varies again between 56 at.% and 62 at.% and Si between 20 at.% and 22 at.%. The ratio between Si and O is between 2.6 to 2.8, accordingly. It should be noted that using oxygen partial pressure during crystal growth is lower than the ratio between Si and O in case of using flux for crystal growth.

The oxygen concentration varies among $BaCuSi_2O_6$ grown under different conditions. We thus suggest to indexing the synthesized powder samples and the single crystals with an additional index "6±δ" to indicate the oxygen content in these compounds, as it is an important parameter for certain physical and chemical properties.

In Fig. 7 a) and b) we show the structural investigation of $BaCuSi_2O_{6\pm\delta}$ by means of x-ray powder diffraction at room temperature and grown with flux, which is in good agreement with powder diffraction pattern of crystals grown by using oxygen partial pressure. The Bragg reflections demonstrate a good crystallinity, and match with the well-known structure typ $I4_1/acd$ at room temperature.

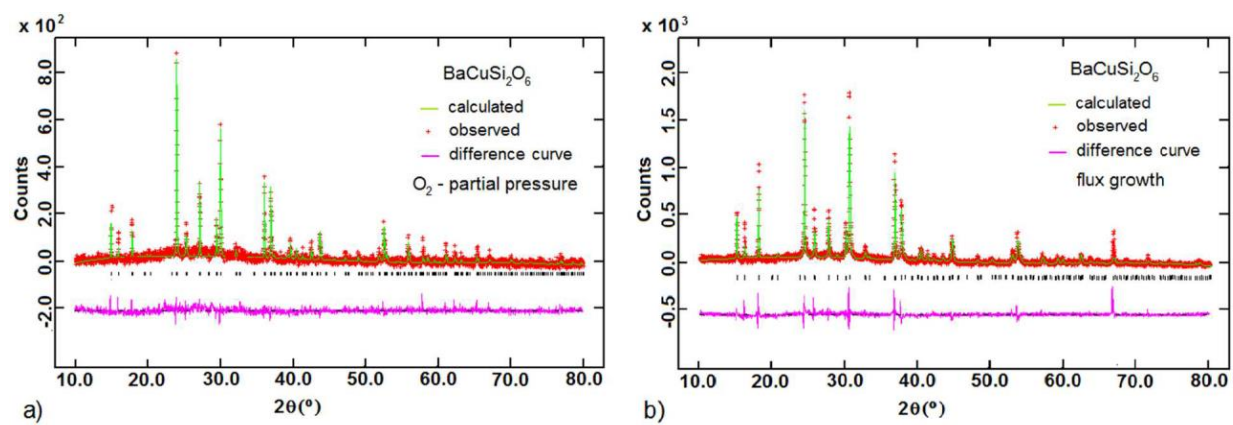

Figure 7: Refined PXRD data, using the GSAS Suite of Rietveld program[18], for BaCuSi$_2$O$_{6\pm\delta}$ grown with a) O$_2$ – partial pressure and b) flux, both showing the tetragonal structure type *I4$_1$/acd*.

The lattice parameters and atomic positions for both samples of BaCuSi$_2$O$_{6\pm\delta}$ are shown in Tab. 5. Both show the tetragonal structure type *I4$_1$/acd*.

Table 5: Lattice parameters and atomic positions at room temperature of BaCuSi$_2$O$_{6\pm\delta}$ grown with flux and by using oxygen pressure, refined from PXRD data

|  | BaCuSi$_2$O$_{6\pm\delta}$ - grown with O$_2$ – partial pressure | | | BaCuSi$_2$O$_{6\pm\delta}$ grown with flux | | |
|---|---|---|---|---|---|---|
| $a = b$ (Å) | 9.969(2) | | | 9.969(2) | | |
| $c$ (Å) | 22.304(7) | | | 22.299(1) | | |
| Atom | x | Y | z | X | y | Z |
| Ba 16e | 0.239(1) | 0 | 0.25 | 0.244(2) | 0 | 0.25 |
| Cu 16d | 0 | 0.25 | 0.066(1) | 0 | 0.25 | 0.062(3) |
| Si 32g | 0.285(1) | 0.763(2) | 0.875(1) | 0.275(1) | 0.766(2) | 0.876(1) |
| O1 32g | 0.177(3) | 0.690(3) | 0.797(2) | 0.211(3) | 0.697(3) | 0.788(1) |
| O2 32g | 0.375(4) | 0.850(4) | 0.871(2) | 0.393(4) | 0.859(3) | 0.883(2) |
| O3 32g | 0.321(3) | 0.762(4) | 0.062(1) | 0.312(2) | 0.761(6) | 0.069(1) |
| $X^2_{red}$ | 1.585 | | | 2.742 | | |
| wRp | 0.1432 | | | 0.1834 | | |
| Rp | 0.1114 | | | 0.1438 | | |

The comparison of the two samples, grown with different methods, shows that the structural symmetry is preserved and that the lattice constants and atomic positions are similar. The data resolution received from diffraction is not sufficient to give an exact statement about the oxygen concentration and variation in BaCuSi$_2$O$_{6\pm\delta}$. However, details about the variation of oxygen in this compound are to be confirmed by high-resolution diffraction data obtained from single crystals. A full investigation by means of synchrotron radiation will be published elsewhere[24].

V. LOW TEMPERATURE STRUCTURE DETERMINATION

The low temperature behavior of the BaCuSi$_2$O$_{6\pm\delta}$ crystals, grown with flux, and the Ba$_{1-x}$Sr$_x$CuSi$_2$O$_{6\pm\delta}$ crystals, grown with oxygen partial pressure have been investigated by powder diffraction measurements. Other sources report a phase transition of BaCuSi$_2$O$_6$ from the tetragonal $I4_1/acd$ structure into the orthorhombic $Ibam$ structure between 110K and 80K[6,11,12]. This phase transition was investigated for the various samples. Fig. 8 summarizes the results from powder diffraction between 140K and 20K, using steps of 20K, together with additional chart at 8K. Worthy of comment is the presence of one reflex combination between 35° and 38° at all investigated temperatures. These reflexes display different intensities at different temperatures. Fig. 8b) shows a detailed view of this angular range. The existence of two phases at 100K and a completed phase transition at 80 K can be clearly seen. In addition, the powder diffraction results of a Sr substituted compound of Ba$_{1-x}$Sr$_x$CuSi$_2$O$_{6\pm\delta}$ grown under oxygen pressure are displayed in Fig 8c) for the angular range 35° to 38°. The lines of the examined Ba$_{0.7}$Sr$_{0.3}$CuSi$_2$O$_{6\pm\delta}$ are slightly broader due to increased disorder in this compound and a reduction of the coherent scattering domains, which might be related to the substitution with Sr. Worthy of comment is the absence of a phase transition at low temperature. In fact, the tetragonal $I4_1/acd$ is preserved down to 20K. This means that the structure has only one type of dimer layers, which have the same distance between the Cu atoms. The underlying mechanism for the suppressed phase transition could be related to a reduced ratio between Si and O ($\pm\delta$). To separate the influence of the substitution with Sr and the influence of the change of the ratio between Si and O ($\pm\delta$), additional investigations of BaCuSi$_2$O$_{6\pm\delta}$, grown under oxygen partial pressure are currently performed. The investigation of the stabilization of the tetragonal structure in will be published elsewhere[25].

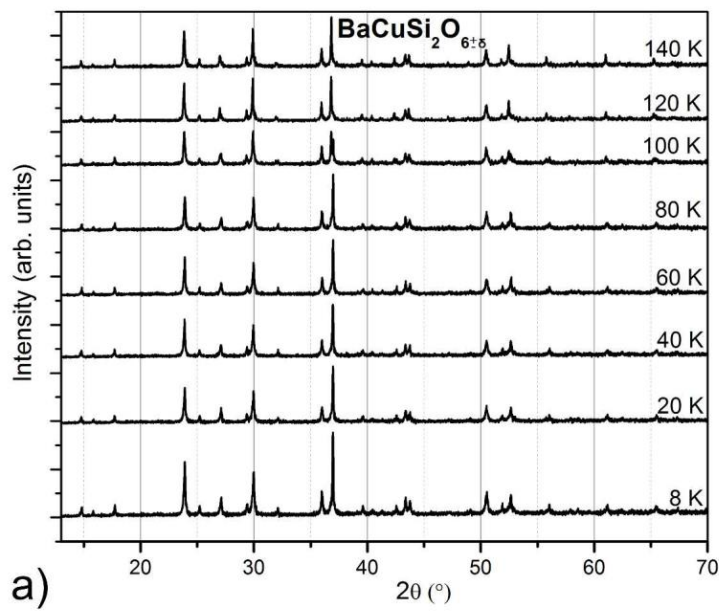

a)

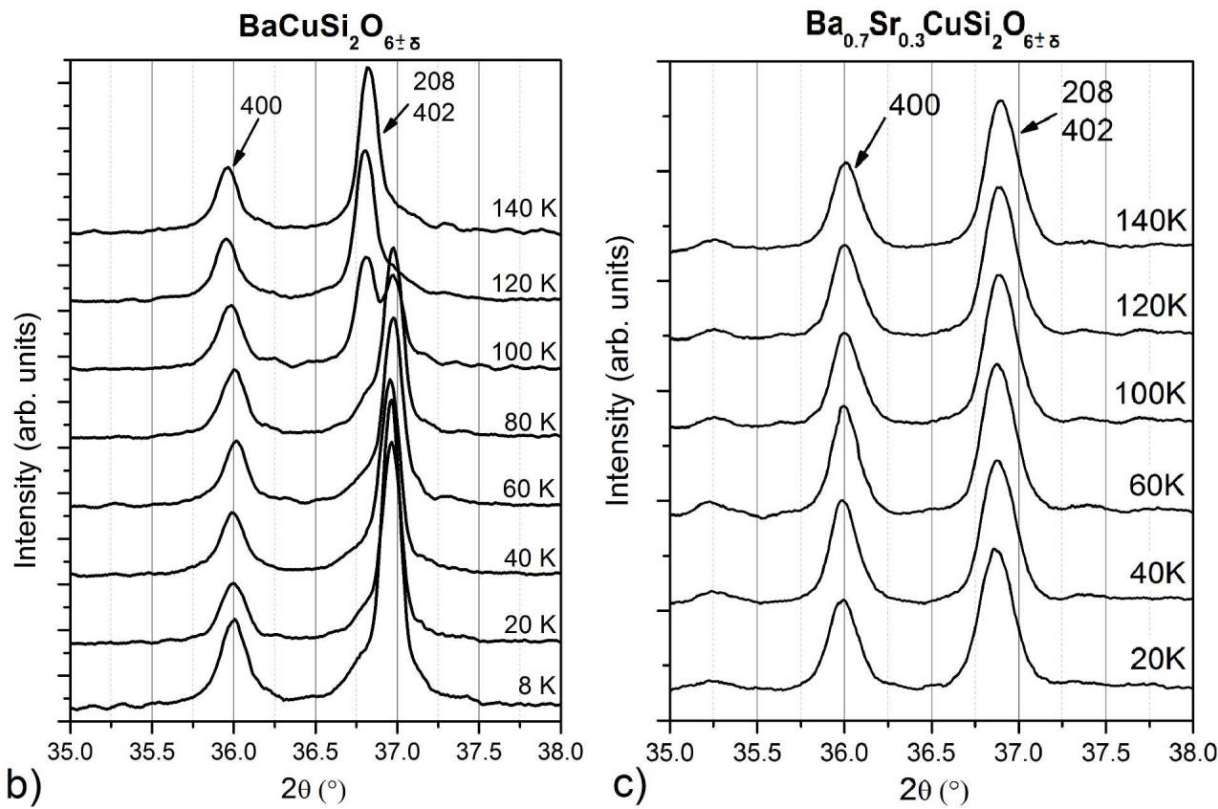

Figure 8: PXRD: a) of BaCuSi$_2$O$_{6\pm\delta}$ at low temperature from 140 K to 8 K; b) of BaCuSi$_2$O$_{6\pm\delta}$ for the selected angle range between 35° and 38° showing the phase transition; and c) of Ba$_{0.7}$Sr$_{0.3}$CuSi$_2$O$_{6\pm\delta}$ for the selected angle range between 35° and 38°, note the absence of a phase transition.

## VI. CONCLUSION AND OUTLOOK

We have shown that the powder synthesis of the layered spin dimer compound BaCuSi$_2$O$_{6\pm\delta}$ with small temperature differences in the high temperature range influences the phase formation process. A flux of 0.5 Na$_2$CO$_3$, 0.2 MgO, 0.1 CaO and 0.2 Al$_2$O$_3$ reduces the sintering temperature of the phase building by 200°C from 1020°C down to 820°C. The DSC examinations show that there is a temperature range for the phase building of BaCuSi$_2$O$_{6\pm\delta}$ (950°C - 1050°C at air), which shifts to highertemperatures by applying oxygen partial pressure (965°C – 1065°C using 0.3 bar oxygen partial pressure).

A substitution of up to 30% Ba by Sr could be carried out successfully. It was found that the sintering temperature for Ba$_{0.7}$Sr$_{0.3}$CuSi$_2$O$_{6\pm\delta}$ powder could be reduced to 1000°C without flux and the structure analysis confirms that the space group $I4_1/acd$ is preserved at room temperature.

Crystal growth of BaCuSi$_2$O$_{6\pm\delta}$ could not only be performed by using flux. The best results for crystal growth of this material was received using oxygen partial pressure of around 1 bar and a temperature of 1150°C.. An oxygen partial pressure is generally required to prevent the chemical decomposition. Using this method with enhanced oxygen partial pressure, also $Ba_{1-x}Sr_xCuSi_2O_{6\pm\delta}$ mixed crystals could be grown, which have the same tetragonal structure type $I4_1/acd$, preserved to low temperatures down to 20K. These crystals have just one type of dimer layers. On first sight, the chemical investigation only reveals slight differences of the ratio of Si and O for $BaCuSi_2O_{6\pm\delta}$ in comparison of the two growth methods. Nevertheless, the ratio of Si and O is very important for preserving the structure. As the EDX analysis of the oxygen content of crystals often fails in the quantitative analysis of light elements, it is advisable to also exercise a wavelength dispersive x-ray (WDX) analysis to determine more precisely the relationship of process parameters and physical properties. The related phase transition has to be further investigated. The variation of the oxygen partial pressure can be assessed as being a control parameter, which also can influence disorder.

The results of the PXRD experiments at low temperatures show that for the series of crystals of $Ba_{1-x}Sr_xCuSi_2O_{6\pm\delta}$ the phase transition from tetragonal to orthorhombic structure is suppressed. To separate the influence of the substitution with Sr and the influence of the change of the ratio between Si and O ($\pm\delta$), additional investigations of $BaCuSi_2O_{6\pm\delta}$, grown under oxygen partial pressure, are currently performed. Such a study is promising to resolve the influence of substitution and the oxygen content for the crystal structure, and, furthermore, contribute to a better understanding of the phase transition process in $Ba_{1-x}Sr_xCuSi_2O_{6\pm\delta}$.

A profound understanding of the magnetic exchange in such compounds and the possibility to create Bose-Einstein condensation requires further investigation of the variation of the oxygen content, and its influence on the Sr-doped system.

* Dr. Natalija van Well, Paul Scherrer Institut, WHGA/344, CH-5232 Villigen PSI, Telephone: +41 56 310 33 87, E-Mail: natalija.van-well@psi.ch

## ACKNOWLEDGMENT

The authors thank D. Chernyshov (SNBL), D. Sheptyakov (PSI) for fruitful discussions, and T. Beier, E. Dautovic for their work on this material during their master theses. This work was supported by the Paul Scherrer Institute and the Deutsche Forschungs-gemeinschaft through SFB/TRR 49 and the research fellowship WE-5803/1-1, and the European Community's Seventh Framework Programme (FP7/2007-2013) under grant agreement n.°290605 (PSI-FELLOW/COFUND).